\documentclass[%
 reprint,
 superscriptaddress,
 nofootinbib,
 amsmath,
 amssymb,
 aps,
 prd,
 a4paper,
 longbibliography
]{revtex4-1}
\pdfoutput=1

\usepackage{graphicx}
\usepackage{dcolumn}
\usepackage{bm}
\usepackage{slashed}
\usepackage[dvipsnames]{xcolor}
\usepackage{physics}
\usepackage{comment}
\usepackage[normalem]{ulem}
\usepackage{hyperref}
\usepackage{url}
\usepackage{tikz}
\usepackage{mathrsfs}

\newcommand{\beq}{\begin{equation} \begin{aligned}}
\newcommand{\eeq}{\end{aligned} \end{equation}}
\renewcommand{\vev}[1]{ \left< {#1} \right> }

\begin{document}

\title{
Gravitational Waves and Nambu-Goldstone Bosons from\\ Infrared-Finite Cosmic Strings
}

\author{Hajime Fukuda}
\email{hfukuda@hep-th.phys.s.u-tokyo.ac.jp}
\affiliation{Department of Physics, The University of Tokyo, Tokyo 113-0033, Japan}

\author{Shuailiang Ge}
\email{shuailiang@uchicago.edu}
\affiliation{Enrico Fermi Institute, The University of Chicago, Chicago, IL 60637, USA}
\affiliation{Department of Physics, Korea Advanced Institute of Science and Technology (KAIST), Daejeon 34141, South Korea}

\author{Keisuke Harigaya}
\email{kharigaya@uchicago.edu}
\affiliation{Enrico Fermi Institute, The University of Chicago, Chicago, IL 60637, USA}
\affiliation{Kavli Institute for Cosmological Physics, Leinweber Institute for Theoretical Physics, Department of Physics, The University of Chicago, Chicago, IL 60637, USA}
\affiliation{Kavli Institute for the Physics and Mathematics of the Universe (WPI),
The University of Tokyo Institutes for Advanced Study,
The University of Tokyo, Kashiwa, Chiba 277-8583, Japan}

\date{\today}

\begin{abstract}
Skyrmion lines and semi-local strings are examples of infrared-finite cosmic strings that arise in theories with spontaneously broken global symmetries. Unlike global $U(1)$ strings, they do not support the long-range Nambu-Goldstone boson (NGB) profile responsible for a logarithmically infrared-divergent string tension. We argue that NGB emission from their low-frequency oscillations is strongly suppressed. Consequently, their gravitational-wave signals may be comparable to those from local strings.
\end{abstract}

\maketitle

\newpage

\noindent {\bf Introduction.}
Symmetry and its spontaneous breaking are central concepts in physics. When a symmetry is broken, the system may admit several energetically equivalent ground states, forming a nontrivial vacuum manifold. If distinct spatial regions relax into different vacua that cannot be smoothly connected everywhere, topological defects can emerge~\cite{Kibble:1976sj,Kibble:1980mv,Vilenkin:1984ib,Zurek:1985qw}.

Various extensions of the Standard Model of particle physics exhibit such symmetry-breaking patterns, and  topological defects may be formed in the early universe. One of the interesting cosmological consequences of topological defects is the emission of gravitational waves (GWs)~\cite{Vilenkin:1981iu,Vachaspati:1984gt}.  Since the characteristic length scale of defect dynamics is much longer than the inverse of the symmetry-breaking scale, we expect that defects emit light particles~\cite{Moore:2001px}. If no light particles other than gravitons couple to the defects, they will lose their energy primarily through the emission of GWs. In particular, the recently discovered stochastic GW background by the pulsar timing array experiments~\cite{NANOGrav:2023gor, NANOGrav:2023hde,NANOGrav:2023hvm} can be explained by primordial GWs from local cosmic strings~\cite{Vilenkin:1982hm,Monin:2008mp,Buchmuller:2019gfy,Buchmuller:2020lbh,Buchmuller:2021mbb,Dunsky:2021tih,Buchmuller:2023aus,An:2023jxf,Lazarides:2023rqf,Bao:2024bws,Chitose:2025cmt,Ingoldby:2025wcl,Bao:2025ori,Asl:2026zpj}.

On the other hand, in theories containing cosmic strings, global continuous symmetries are often spontaneously broken, and the associated Nambu-Goldstone bosons (NGBs) may couple to the strings. Because NGBs are massless, they can in principle be emitted efficiently from cosmic strings. If so, GW emission is suppressed, and the resulting signal may be too weak either to explain the pulsar timing array observations or to be detected in future searches. Indeed, global $U(1)$ strings can radiate NGBs efficiently because the NGB field has a long-range profile, which leads to a logarithmically infrared-divergent string tension.

In this Letter, we study NGB emission from skyrmion lines and semi-local strings, which are examples of infrared-finite cosmic strings. Their energy density is localized and their oscillations do not excite NGBs at long distances. We show that NGB emission from their low-frequency oscillations is therefore strongly suppressed. As a result, their GW signals may be comparable to those of local strings, broadening the class of symmetry-breaking patterns that can efficiently generate observable primordial GWs.

\vspace{0.2cm}
\noindent {\bf Global $U(1)$ string.}
Before studying skyrmion lines and semi-local strings, we review the reason why global $U(1)$ strings efficiently emit NGBs. When a global $U(1)$ symmetry is spontaneously broken by a complex scalar field $\phi$ with a vacuum expectation value $v$, the following string configuration can exist,
\begin{equation}
    \phi = v \times g(r) e^{i\theta}.
\end{equation}
Here we use cylindrical coordinates $(r,\theta,z)$ and place the string along the $z$ axis. $g(r)$ is determined by the equation of motion of $\phi$ with boundary conditions $g(0)=0$ and $g(\infty)=1$. The gradient energy of the string coming from the $\theta$ direction has a logarithmic IR divergence. This means that the NGB mode, corresponding to the angular direction of $\phi$, carries appreciable gradient energy even far from the string core. In the cosmological context, the divergence is cut off by the distance between strings.

When the string oscillates, the NGB configuration is perturbed even at large distances, leading to efficient NGB emission. This is analogous to the emission of electromagnetic waves from a particle in $D=2+1$ or a uniformly charged line in $D=3+1$, where the electromagnetic field energy also exhibits a logarithmic infrared divergence. Note that this emission is not suppressed by the NGB energy, despite the usual derivative suppression of NGB interactions~\cite{Weinberg:1968de}. Intuitively, this is because a substantial gradient energy is already stored in the NGB configuration around the string, and the oscillation excites this background rather than creating NGBs from the vacuum. Note that the low-energy theorem of NGBs is applicable only to small excitation of NGBs around the vacuum.

The efficient emission by analogy with electromagnetism becomes clearer by taking a dual of the theory, where the NGB becomes the gauge field of a one-form gauge symmetry under which the string has a unit charge~\cite{Witten:1985fp,Vilenkin:1986ku,Davis:1988rw}. Then the NGB emission can be interpreted as  gauge field emission from a (magnetically) charged object, whose charge corresponds to the winding number of the string that is topological and can be measured from infinity by accumulating the change in the $\theta$ direction, in analogy with the measurement of electromagnetic fields.

As we will see, the energy densities of skyrmion lines and semi-local strings are localized. Their motion therefore does not induce long-range oscillations of the NGB field.
One can also construct an effective theory in which the string is treated as a line defect and the NGB coupling involves derivatives, as expected from the low-energy theorem.

\vspace{0.2cm}
\noindent {\bf Skyrmion line.}
Skyrmion lines, whose cross sections are known as baby skyrmions\,\cite{Skyrme:1961vq,Piette:1994ug,Piette:1994mh}, are topological defects associated with the breaking of a global symmetry $G$ to $H$ such that $\pi_2(G/H) = {\bf Z}$. To be concrete, we consider $G=SU(2)$ and $H=U(1)$. There are two NGBs that form a $U(1)$ charged complex scalar field $\chi$.

Let $\phi^a$ be a scalar field in the adjoint representation of $SU(2)$ with a wine-bottle potential with a minimum at $\phi^a \phi^a=v^2$. We assume that the potential is sufficiently steep that the radial mode of $\phi^a$ can be integrated out. $\phi^a$ then describes a non-linear sigma model with a target space $G/H \cong S^2$. The leading-order Lagrangian does not stabilize the skyrmion line because of Derrick’s theorem~\cite{Derrick:1964ww}; the tension of skyrmion lines from $n$-th derivative terms scales as $R^{2-n}$, where $R$ is the radius of the skyrmion line. There are several ways to stabilize the radius. For example, if we include an $n=4$ term with a negative coefficient and an $n=6$ term with a positive coefficient, the balance of the two terms may stabilize the radius. Alternatively, we may introduce a small explicit breaking of $SU(2)$ by $n=0$ term, which favors configurations that stay as close as possible to the vacuum and hence prefer smaller radii, so if the $n=4$ term favors large radii, they may stabilize the radius as well\,\cite{Piette:1994ug,Piette:1994mh}.

To construct the solution, we first define
\begin{equation}
    \Phi = \sigma^a \phi^a
\end{equation}
and place the string along the $z$ axis. The skyrmion line configuration is given by
\begin{align}\label{eq:skyrmion_ansatz}
    \Phi_S({\bf x}) =& v V \sigma^3 V^\dag, \\
    V =& {\rm exp} \left( i \frac{\sigma^i x_\perp^i}{2r} f(r)\right) =
    \begin{pmatrix}
    {\rm cos}\frac{f}{2} & i e^{-i\theta}{\rm sin}\frac{f}{2} \\
    i e^{i\theta}{\rm sin}\frac{f}{2} & {\rm cos}\frac{f}{2}
    \end{pmatrix}, \nonumber
\end{align}
where $x_\perp^i = (x,y) = (r\cos\theta,r\sin\theta)$ and $f(r)$ is a function with a boundary condition $f(\infty)=0$ and $f(0)=\pi$. To the leading order in derivative expansion, it satisfies
\begin{equation}
\label{eq:feq}
    f'' + \frac{1}{r}f' - \frac{1}{2r^2}{\rm sin}(2f)=0,
\end{equation}
whose solution is given by $f(r) = 2 \arctan(r_0/r)$. Here $r_0$ is the radius of the skyrmion line that is not fixed by the leading order Lagrangian, but it is fixed by higher derivative terms or explicit breaking of $SU(2)$, as discussed above.

The leading-order contribution to the string tension is given by
\begin{equation}
\label{eq:mu}
   \mu =  \int d^2 x_\perp \frac{1}{4} {\rm tr}\left(\partial_i \Phi_S \partial_i \Phi_S\right).
\end{equation}
Unlike for global $U(1)$ strings, the tension in Eq.~\eqref{eq:mu} is finite; since $f\propto r^{-1}$ as $r \rightarrow \infty$, the NGB excitation is suppressed far from the string core. The oscillation of NGB fields induced by the oscillation of the string is suppressed, which corresponds to suppressed emission of NGB fields.

\vspace{0.2cm}
\noindent {\bf Semi-local string.}
Semi-local strings are defects associated with the breaking of both global and gauge symmetries~\cite{Vachaspati:1991dz,Preskill:1992bf}. Suppose a symmetry $G$ is spontaneously broken to $H$, and their subgroups $G_{\rm gauge}$ and $H_{\rm gauge}$ are gauged. Even if $\pi_1(G/H)$ is trivial, if $\pi_1(G_{\rm gauge}/H_{\rm gauge})$ is non-trivial, topological defects may arise, and they are called semi-local strings. To be concrete, we consider $G = SU(2)\times U(1)_{\rm gauge}$, where $U(1)_{\rm gauge}$ is gauged, and the unbroken subgroup is the diagonal subgroup, $H= U(1)$, which is a global symmetry. There are again two NGBs that form a complex scalar field $\chi$ charged under the unbroken $U(1)$. This type of setup naturally arises when $U(1)_{\rm gauge}$ is spontaneously broken by QCD-like dynamics (see e.g.~\cite{Harigaya:2016rwr,Co:2016akw,Contino:2020god,Ibe:2021gil,Fukuda:2026bdn}).

The semi-local string has a close connection to the skyrmion line and Abrikosov-Nielsen-Olesen (ANO) vortex~\cite{Hindmarsh:1991jq,Hindmarsh:1992yy,Preskill:1992bf}. One possible configuration is an ANO vortex, in which the gauged $U(1)_{\rm gauge}$ winds around the vortex and the Higgs field value vanishes at the core. There can also be a skyrmion-like configuration with the same asymptotic winding but a nonvanishing Higgs field value at the core. To see that this configuration may be topologically stable, let us consider a low-energy limit where the radial direction of the Higgs field and the massive gauge boson are integrated out. The low-energy effective theory is a non-linear sigma model with a target space $S^3/U(1)_{\rm gauge} \cong S^2$, which contains a skyrmion line configuration associated with the homotopy group $\pi_2(S^2) = {\bf Z}$. (Mathematically, the correspondence of these two configurations stems from the long exact sequence of homotopy groups for the Hopf fibration.)

Whether a stable solution exists depends on the details of the theory. In a perturbative model where the Higgs quartic coupling is $\lambda$ and the $U(1)$ gauge coupling is $e$, skyrmion line-like configurations are unstable; the scalar field configuration is scale-invariant, but these configurations have a magnetic flux of $U(1)_{\rm gauge}$, which favors a larger radius. Thus, only when the ANO vortex has a smaller energy than the skyrmion line, the semi-local string can be stable and topological defects are formed. The condition determining which of the two configurations has the lower energy is given by the ratio of the quartic coupling of the Higgs field and the $U(1)_{\rm gauge}$ coupling, $\beta \equiv \lambda/e^2$. The ANO string has a smaller energy for $\beta < 1$, while the skyrmion line has a smaller energy for $\beta > 1$. When $\beta = 1$, the semi-local string saturates the Bogomol'nyi-Prasad-Sommerfield (BPS) bound and a modulus of the solution interpolates the ANO and skyrmion-like line solutions; we may even embed the semi-local string into $\mathcal{N}=2$ supersymmetric theory\,\cite{Tong:2005un}. The details of the Lagrangian and the string profile are reviewed in Appendix~\ref{app:semi-local}.

In non-perturbative models, higher-derivative terms in the non-linear sigma model can stabilize the skyrmion line-like solution. (It is in principle possible that an ANO solution exists, but we cannot use the non-linear sigma model to analyze the stability of ANO configuration.)

So far we have considered the case of exact $SU(2)$ symmetry, but explicit breaking can allow skyrmion line-like solutions over a broader range of parameters; a non-zero potential of the NGB fields favors a smaller radius of the skyrmion line-like solution, which  may balance the effect of the magnetic flux of $U(1)_{\rm gauge}$ that favors a larger radius, both in perturbative models with $\beta >1$ and non-perturbative models.

\vspace{0.2cm}
\noindent {\bf Effective coupling between NGBs and strings.}
We introduce the NGB fluctuations around the skyrmion solution as
\begin{equation}
    \Phi(x) = U \Phi_S U^\dag,~~U= {\rm exp}\left(i \sum_{i=1,2}\sigma^i \frac{\Pi^i(x)}{2v}\right).
\end{equation}
Expanding the Lagrangian to the second order in $ \chi \equiv (\Pi^1 + i\Pi^2)/\sqrt{2}$ and approximating the skyrmion line as a point-like particle, we obtain (see Appendix~\ref{app:semi-local coupling} for the derivation)
\begin{align}
\label{eq:Leff}
\mathcal{L}_{\rm eff}
\simeq& 
-\frac{1}{v^2}\left(
N_s \partial_{\mu}\chi\partial^{\mu}\chi^*+
N_ai\epsilon_{ij}\partial_i\chi^{*}\partial_j\chi
\right) \nonumber \\
&\times \delta^{(2)}(\boldsymbol{x}_{\perp} - \boldsymbol{X}(t,z)),
\end{align}
where
\begin{align}
\label{eq:N}
N_s \simeq N_a = & \pi v^2 
\int dr r
\sin^2 f(r),
\end{align}
$i=x,y$, and  $\boldsymbol{X}(t,z)$ is the position of the skyrmion line in the $xy$ plane. The NGBs enter only through derivatives. There are no terms linear in the NGB field: since $\chi$ is charged under the unbroken $U(1)$ whereas the skyrmion line is neutral, such terms are forbidden by symmetry.

$N_{s}$ and $N_a$ have logarithmic IR divergences since $f\propto r^{-1}$ at $r \rightarrow \infty$. The IR divergence should be cut off around the wavelength of the NGB, $k^{-1}$, which is the typical spatial size of the wavefunction of the final states. When the wavelength is much longer than the size of the skyrmion core, $N_{s,a} \sim 4\pi v^2 r_0^2 {\rm ln}[1/(k r_0)]$, for which the contribution to Eq.~\eqref{eq:N} from, say, $r < 0.1k^{-1}$, is much larger than that from $r > 0.1k^{-1}$. Thus, the point-like particle approximation is valid.

The coupling between NGBs and a semi-local string is derived in the Appendix and takes the same form as Eq.~\eqref{eq:Leff}; see Appendix~\ref{app:semi-local coupling}. For ANO-like strings, $N_s\sim 1/\lambda $ and  $N_a\sim 1/e^2$ without logarithmic enhancement, because the NGB is not excited anywhere on or outside the string. For skyrmion line-like strings, $N_s \sim 1/\lambda$ while $N_a$ is logarithmically enhanced. 

\vspace{0.2cm}
\noindent {\bf NGB emission rate.}
When the string oscillates, NGBs may be emitted in pairs via the interaction in Eq.~\eqref{eq:Leff}. To demonstrate suppressed emission of NGBs, we consider a straight string that oscillates with an amplitude $X_0$ and a frequency $\omega$, $\boldsymbol{X}(t,z) = X_0 {\rm cos}(\omega t) \hat{n}_x$. The energy of the NGBs emitted from the string per length and per time is, to the leading order in $X_0 \omega$,
\begin{align}
\label{eq:dE/dtdL}
    \frac{d^2E}{dt dL} \simeq
    \frac{ N^2X_0^2 \omega^9}{12288 \pi^2 v^4},~~~ N^2 \equiv N_s^2 + \frac{1}{3}N_a^2,
\end{align}
where we assume $m_\chi \ll \omega/2$. The radiation power is proportional to $X_0^2$ and to the square of the coupling in Eq.~\eqref{eq:Leff}, and the dependence on $\omega$ can be determined by dimensional analysis. The detailed computation is provided in Appendix~\ref{app:emission rate}. For $m_\chi > \omega/2$, the emission rate is lower.

Our computation provides only a crude estimate of the emission rate in realistic string configurations, where $X_0\omega = \mathcal{O}(1)$ and the string motion is generically not that of a simple harmonic oscillator. Note that ${\bf X}(t,z) = X_0 {\rm cos}(\omega t) \hat{n}_x$ is not a solution of the equation of motion of a freely moving string and requires an external force. The estimated rate can be understood as that from an approximately straight part of the string network that is forced to oscillate by the tension from other parts of the network. The emission rate can depend on the details of the configuration. For example, if we instead consider ${\bf X}(t,z) = X_0 {\rm cos}(\omega t) {\rm cos}(\omega z) \hat{n}_x$, which is a solution of a free string, we find that the rate is proportional to $X_0^4 \omega^{11}/v^4$. The $O(X_0^2)$ contribution vanishes due to momentum conservation in the $z$ direction. In realistic cosmological setups, strings will not be straight and momentum conservation will not hold, so the $O(X_0^2)$ contribution may not vanish. Below, we will take $X_0\sim \omega^{-1}$ and thus expect that such details do not alter the conclusion that the NGB emission rate is suppressed.

\vspace{0.2cm}
\noindent {\bf Cosmological implications.}
In the early universe, strings can be produced by symmetry-breaking phase transition. Assuming that string reconnection is efficient and the friction on the strings is negligible, the string network enters the scaling regime, where there are $O(1)$ strings of horizon-scale length per horizon volume~\cite{Bennett:1987vf,Martins:1996jp}. String reconnection produces loops~\cite{Vanchurin:2005pa,Blanco-Pillado:2011egf}, which oscillate and shrink while emitting particles.

For a loop with a radius $\ell$, the typical frequency and the amplitude of the oscillation are $O(\ell^{-1})$ and $O(\ell)$, respectively. To estimate the radiation power into NGBs, we take $X_0= \omega^{-1}= \ell$ in Eq.~\eqref{eq:dE/dtdL} to obtain
\begin{equation}
    P_{\rm NGB} = \frac{N^2}{12288 \pi^2 v^4 \ell^6},
\end{equation}
which is much smaller than that of global $U(1)$ strings $\sim v^2$~\cite{Battye:1993jv}. The radiation power into GWs is~\cite{Blanco-Pillado:2017oxo}
\begin{equation}
    P_{\rm GW} \simeq \Gamma \frac{\mu^2}{8 \pi M_{\rm pl}^2},
\end{equation}
where $\Gamma \sim 50$ is a dimensionless constant. The radiation into GWs dominates if
\begin{equation}
    \ell > 0.1 \frac{1}{v} \left( \frac{N M_{\rm pl}}{v} \frac{\pi v^2}{\mu} \right)^{1/3}.
\end{equation}

The typical initial size of the loops is the horizon size when they are produced. Assuming radiation domination, the loops predominantly decay into GWs if they are produced at temperatures satisfying
\begin{equation}
    T  < 2  v  N^{-1/6} \left( \frac{M_{\rm pl}^2}{v^2} \frac{\mu}{\pi v^2}\right)^{1/6}. 
\end{equation}
We therefore expect that string loops produced after the phase transition and in the scaling regime predominantly decay into GWs. A non-zero mass of the NGB by explicit global symmetry breaking can further suppress the NGB emission.

The resulting GW abundance is then expected to be similar to that from local strings and is given by~\cite{Auclair:2019wcv}
\begin{equation}
\label{eq:GW}
    \Omega_{\rm GW}h^2 \simeq 3\times 10^{-9} \frac{v}{10^{15}~{\rm GeV}}
\end{equation}
at high frequencies. The abundance is much larger than that from global $U(1)$ strings, which is suppressed due to efficient NGB emission.

\vspace{0.2cm}
\noindent {\bf Discussion.}
In this Letter, we have shown analytically that NGB emission from oscillating skyrmion lines and semi-local strings is parametrically suppressed. These defects provide examples of infrared-finite cosmic strings, in contrast to global $U(1)$ strings whose long-range NGB profiles lead to logarithmically infrared-divergent tensions and efficient NGB radiation. Although our computation focuses on skyrmion lines and semi-local strings, we expect that the suppressed NGB emission is a generic feature of infrared-finite cosmic strings. Our result therefore broadens the class of symmetry-breaking patterns that can efficiently generate observable primordial GWs.

We have demonstrated the suppressed interaction of NGBs with strings and explicitly computed the emission rate of a pair of NGBs. In some theories and configurations, there can be $U(1)$ moduli associated with the unbroken $U(1)$, which can affect the emission of NGBs. For ANO-like semi-local strings, only the scalar component neutral under the unbroken $U(1)$ is excited around the string, so no $U(1)$ moduli exist. For BPS semi-local strings, a $U(1)$ modulus and a size modulus appear, which may be understood as supersymmetric partners of each other\,\cite{Leese:1992fn,Abraham:1992hv,Tong:2005un}.  Although the moduli metric is divergent and non-renormalizable for isolated strings (see Appendix~\ref{app:semi-local_BPS}), which reflects the fact that we need a steady flux of incoming and outgoing NGBs to keep the $U(1)$ charge around strings and otherwise the $U(1)$ charge is radiated away, multi-string configurations may have a normalizable $U(1)$ moduli~\cite{Leese:1992fn,Abraham:1992hv}. For isolated skyrmion-line like solutions, if the stabilization does not involve explicit $SU(2)$ breaking, a $U(1)$ modulus exists but is non-normalizable (see Appendix~\ref{app:Skyrmion_charge}.) If the stabilization involves explicit $SU(2)$ breaking, a normalizable $U(1)$ moduli exists~\cite{Piette:1994mh}. For both stabilization mechanisms, for multi-string configurations, there might be normalizable $U(1)$ moduli, although we are not aware of a systematic study of the moduli of skyrmion-line like solutions.

Unlike skyrmions and monopoles in $D = 3+1$, normalizable $U(1)$ moduli, if exist, are massless excitations living on the string, and can be excited even if the oscillation frequency $\omega$ is much below the symmetry-breaking scale $v$. Since these moduli also carry the unbroken $U(1)$ charge, there can be processes where one NGB is emitted while one $U(1)$ modulus is excited on the string. Because the wave function of the $U(1)$ modulus is localized on the string, the rate of such a process can be larger than that of the emission of a pair of NGBs. However, because of the suppression by the derivative on the NGB, the relevant interaction term in the Lagrangian is still proportional to $1/v$, and the energy of NGBs emitted from the string per length and per time is expected to be proportional to $X_0^2 \omega^7/v^2$. The resultant radiation power from a loop is proportional to $1/(v^2 \ell^4)$, and the emission of GWs still dominates.

Even if the NGB emission is suppressed, a precise relation between the GW abundance and the symmetry-breaking scale will require dedicated numerical simulations of skyrmion lines and semi-local strings. In particular, in Eq.~\eqref{eq:GW}, we assume that the production of loops is as efficient as in local $U(1)$ string cases, but this efficiency should be quantified by lattice simulations. 

At first sight, our result appears inconsistent with the lattice simulation of semi-local strings in~\cite{Kanda:2025hgi}, which finds an NGB number density of order $v^2 H$, where $H$ is the Hubble expansion rate. This number density seems consistent with efficient NGB emissions from strings that follow the scaling law. However, the observed number density can be also explained by the initial fluctuations of NGBs. With the randomized initial conditions of~\cite{Kanda:2025hgi}, there are NGB fluctuations with $\vev{|\chi|^2} \sim v^2$ outside the horizon. As they enter the horizon, they begin to oscillate, producing a number density $\sim v^2 H$.
Furthermore, the semi-local string systems include global monopoles as endpoints of string segments~\cite{Achucarro:1999it}, and the NGBs observed in~\cite{Kanda:2025hgi}  might originate from the global monopoles~\cite{Kanda:2026oon}.  
To avoid possible contamination from the pre-existing NGB fluctuations and emissions from global monopoles, it would be useful to simulate an isolated string loop without fluctuations outside its core and measure the evolution of the number density of NGBs. Inclusion of the cosmic expansion will not be crucial for this simulation.

For the numerical computation of the NGB emission rate from loops, we note that the loop lifetime remains under debate even for local $U(1)$ strings. Some field-theoretic simulations find that loops disappear on a timescale $\sim \ell$~\cite{Hindmarsh:2008dw,Hindmarsh:2021mnl}. The physical origin of this short lifetime in these simulations has not been identified, although~\cite{Hindmarsh:2021mnl}
suggests that the efficient energy loss may arise from massive excitations on the string. While it remains unclear whether such a short loop lifetime persists in the actual early universe, isolating the possible effect of massive modes from NGB emission requires simulations of skyrmion lines and semi-local strings in setups where local $U(1)$ string loops are long-lived. For example, the ``artificial string configurations'' introduced in~\cite{Hindmarsh:2021mnl}, in which massive excitations are suppressed, may provide a useful framework.

\vspace{0.3cm}
\noindent {\it Acknowledgment.}
The authors thank Yukihiro Kanda for discussion on the possible NGB emission from global monopoles and Dam T.~Son for discussion on $U(1)$ moduli. 
The authors acknowledge the use of OpenAI’s ChatGPT and Anthropic’s Claude for assistance with language polishing and analytical computations.
This work was supported by the Department of Energy grants DE-SC0009924 [KH]; the JSPS KAKENHI Grant Nos.\ 25H00638 [HF], 26H00403 [HF] and 26K17131 [HF]; and the World Premier International Research Center Initiative (WPI), MEXT, Japan (Kavli IPMU) [KH].
In this research work, HF used the UTokyo Azure (\url{https://utelecon.adm.u-tokyo.ac.jp/en/research_computing/utokyo_azure/}).


\appendix

\section{Skyrmion lines}

\subsection{$U(1)$ charge on skyrmion lines}
\label{app:Skyrmion_charge} 

In this appendix we argue that isolated skyrmion lines without explicit breaking of global symmetry cannot have $U(1)$ charges localized on the lines. Let us attempt to introduce a non-zero $U(1)$ charge via
\begin{equation}
    \Phi = e^{i \alpha (t) \sigma^3/2} \Phi_S e^{-i  \alpha (t)\sigma^3/2}.
\end{equation}
The tension of the string is then
\begin{equation}
    \mu = v^2 \int d^2 x_\perp \left( \frac{1}{2} f'^2 + \frac{1}{2}\left(\frac{1}{ r^2} +  \dot{\alpha}^2 \right) {\rm sin}^2 f \right),
\end{equation}
where the primes and dots denote the derivatives with respect to $r$ and $t$, respectively.
The tension would exhibit a logarithmic divergence if $f$ is approximated by the solution of Eq.~\eqref{eq:feq}. In other words, the metric of the modulus $\alpha$ is logarithmically divergent.
It is expected that $f$ is significantly affected in the presence of non-zero $U(1)$ charges. 

The equation of motion of $\alpha$ is
\begin{align}
  \left( \frac{\partial^2}{\partial t^2} - \frac{\partial^2}{\partial z^2}\right) \alpha=0, 
\end{align}
where we include a possible dependence of $\alpha$ on $z$ to be more generic. Oscillating solutions of $\alpha \propto {\rm cos}(\omega(t\pm z))$ do not correspond to non-zero $U(1)$ charges on the string. We therefore consider a solution with $\alpha  = \omega t$.
The equation of motion for $f$ is
\begin{align}
\label{eq:eom with charge}
  f'' + \frac{1}{r}f' - \frac{1}{2}\left(\frac{1}{r^2} -  \omega^2
   \right)   {\rm sin}2f=0.
\end{align}
Since $f\rightarrow 0$ as $r\rightarrow \infty$, we can approximate $\sin 2f$ by $2f$ at large distances. Also, for $r\gg \omega^{-1}$, the term $\propto r^{-2}$ is negligible. Linearized Eq.~\eqref{eq:eom with charge} for $r\gg \omega^{-1}$  has a general solution
\beq
f(r) \simeq c_1 H_0^{(1)}(\omega r) + c_2 H_0^{(2)}(\omega r),
\eeq
where $H_0^{(1,2)}$ are Hankel functions of the first and second kind.  Their large distance behavior is $H_0^{(1)}(\omega r) \simeq \sqrt{2/(\pi\omega r)}{\rm e}^{i(\omega r - \pi/4)}$ and $H_0^{(2)}(\omega r) \simeq \sqrt{2/(\pi\omega r)}{\rm e}^{-i(\omega r + \pi/4)}$ for $\omega r\gg 1$. Therefore, if $\omega \neq 0$, the solution for $f$ at $r \rightarrow \infty$ decreases in proportion to $r^{-1/2}$.
Combined with the time-dependence of the asymptotic NGB fields, the two Hankel components describe outgoing and incoming waves. Since $f(r)$ is real, we have $c_2 = i c_1^{*}$, corresponding to a standing-wave superposition.

The $U(1)$ charge per length is given by
\begin{equation}
\label{eq:q}
    q =  v^2 \dot{\alpha} \int d^2 x_\perp {\rm sin}^2 f.
\end{equation}
For $f\propto r^{-1/2}$, the integration in Eq.~\eqref{eq:q} is dominated at large $r$, which means that the charge is not localized around the skyrmion line. Rather, this solution corresponds to incoming and outgoing waves of NGBs as described above.
The total energy per unit length diverges linearly, which is analogous to the linear divergence of plane-wave solutions in $D=1+1$. (In physical situations, radiation of NGBs occurs through wave packets that have a finite energy.)

Ref.~\cite{Piette:1994mh} numerically solved the dynamics of skyrmion lines with $U(1)$ charges, but with explicit $SU(2)$ breaking without explicit $U(1)$ breaking, which gives a non-zero NGB mass $m_\chi$. For $|\dot{\alpha}|> m_\chi$, which should mimic our setup without $m_\chi$, the $U(1)$ charges around the skyrmion line are quickly radiated away, confirming our argument. For $|\dot{\alpha}|< m_\chi$, on the other hand, the $U(1)$ charges continue to be localized around the skyrmion line. In fact, when $m_\chi$ is included in Eq.~\eqref{eq:eom with charge}, $f$ exponentially decreases for $r > m_{\chi}^{-1}$, and hence the integration in Eq.~\eqref{eq:q} is dominated by $r < m_{\chi}^{-1}$.

We conclude that there is no stable configuration with $U(1)$ charges localized around skyrmion lines, unless explicit global symmetry breaking is introduced.

\subsection{Coupling of skyrmion lines to NGBs}
\label{app:skyrmion coupling}

To derive the coupling of skyrmion lines to NGBs, we consider the NGB modes on top of the skyrmion-line solution $\Phi_S$,
\begin{equation}
    \Phi(x) = U \Phi_S U^\dag,~~U= {\rm exp}\left(i \sum_{a=1,2}\sigma^a \frac{\Pi^a(x)}{2v}\right).
\end{equation}
Substituting it into the Lagrangian $\mathcal{L} = \frac{1}{4} {\rm tr}(\partial_{\mu}\Phi \partial^{\mu}\Phi)$, we obtain the interaction terms between the skyrmion line and NGBs,
\beq\label{eq:NGB_skyrmionLine}
\mathcal{L}_{\rm int} 
&
=
\frac{1}{v}(\partial_{\mu}\chi j_{-}^{\mu} + \partial_{\mu}\chi^* j_{+}^{\mu}) \\
&-
\frac{i}{2v^2}(\chi^*\partial_{\mu}\chi - \chi\partial_{\mu}\chi^*)j_3^{\mu} \\
&-
\frac{1}{4 v^2}
{\rm tr}([T_a,\Phi_S][T_b, \Phi_S])
\partial_{\mu}\Pi_a\partial^{\mu}\Pi_b \\
& + \mathcal{O}(\Pi^3)
,~~~(a,b=1,2)
\eeq
where $\chi \equiv (\Pi_1 + i\Pi_2)/\sqrt{2}$, $T_a \equiv \sigma_a/2$, $T_{\pm}\equiv (T_1\pm iT_2)/\sqrt{2}$, and $j_a$ are the conserved currents corresponding to the generators $T_a$, 
\beq
j_a^{\mu} = \frac{i}{2}{\rm tr}(T_a[\Phi_S, \partial^{\mu}\Phi_S])
,~~~
\partial_{\mu}j_a^{\mu} = 0,
\eeq
and $j_{\pm}^{\mu} \equiv (j_1^{\mu}\pm ij_2^{\mu})/\sqrt{2}$. Because $\partial_{\mu}j_{\pm}^{\mu} = 0$, the first line in Eq.~\eqref{eq:NGB_skyrmionLine} is a total derivative and can be dropped.

With the radially symmetric ansatz in Eq.~\eqref{eq:skyrmion_ansatz} substituted into $j_3^{\mu}$,
\beq
j_3^i = -v^2 \sin^2 f(r) \partial^i \theta.
\eeq
Defining $H(r) = \int_r^{\infty} d\rho~\sin^2f(\rho)/\rho$,
\beq
\partial_j H = H'(r) \frac{r_j}{r}
,~~~
\epsilon_{ij}\partial_j H
=
\sin^2 f(r) \partial_i \theta.
\eeq
Therefore,
\beq
j_3^i =  -v^2 \epsilon_{ij} \partial^j H(r).
\eeq
For nearly straight skyrmion lines that slowly oscillate,
$\partial_{0,z}$ acting on $\Phi_S$ are negligible and thus $j_3^{0,z} \simeq 0$. Consequently, with the boundary term vanishing, we can express the second line of Eq.~\eqref{eq:NGB_skyrmionLine} as $-iH(r) \epsilon_{ij}\partial_i \chi^*\partial_j\chi$.
Also, using $\Phi_S = \phi_a\sigma_a$, the third line of Eq.~\eqref{eq:NGB_skyrmionLine} can be written as
$
(\frac{1}{2}
\partial_{\mu}\Pi_a\partial^{\mu}\Pi_a
-
\frac{1}{2}\frac{\phi_a\phi_b}{v^2}
\partial_{\mu}\Pi_a\partial^{\mu}\Pi_b)
$,
where $\phi_1  = - v\sin f \sin\theta$ and $\phi_2 = v\sin f \cos\theta$.

In summary, Eq.~\eqref{eq:NGB_skyrmionLine} can be rewritten as (with the free kinetic term $\partial_{\mu}\chi \partial\chi^*$ subtracted),
\beq
\mathcal{L}_{\rm int}
&=
-iH(r) \epsilon_{ij}\partial_i \chi^*\partial_j\chi 
-\frac{\sin^2 f(r)}{2}
\partial_{\mu}\chi\partial^\mu\chi^* \\
&+
\frac{\sin^2 f(r)}{4}
({\rm e}^{-2i\theta}\partial_{\mu}\chi\partial^{\mu}\chi
+
{\rm e}^{2i\theta}
\partial_{\mu}\chi^*\partial^{\mu}\chi^*
).
\eeq

Approximating the skyrmion line as a thin line defect, the Lagrangian becomes
\beq
\mathcal{L}_{\rm eff}
\simeq& 
-\left(
\frac{N_a}{v^2}i\epsilon_{ij}\partial_i\chi^{*}\partial_j\chi
+
\frac{N_s}{v^2} \partial_{\mu}\chi\partial^{\mu}\chi^*
\right) \\
&\times \delta^{(2)}(\boldsymbol{x}_{\perp} - \boldsymbol{X}(t,z)),
\eeq
where
\beq
N_a\equiv v^2 \int d^2 x_\perp 
 H(r),~
N_s \equiv 
\frac{v^2}{2}\int d^2 x_\perp \sin^2 f(r).
\eeq
The subscripts `$a$' and `$s$' represent the antisymmetric and symmetric term, respectively. 
$\boldsymbol{X}(t,z)$ is the position of the skyrmion line on the $xy$ plane.
The terms $\partial\chi\partial\chi$ and $\partial\chi^*\partial\chi^*$ vanish in this effective description, since $\int d^2 x \sin^2 f {\rm e}^{\pm 2i\theta} = 0$.

As discussed in the main text,
$N_s$ has an IR logarithmic divergence because $f(r)\rightarrow r^{-1}$ as $r\rightarrow \infty$. In addition, 
\beq
N_a &= 2\pi v^2 \int_{r=0}^{r=r_{\rm cut}} dr ~ rH(r)
\\
&=2\pi v^2 \int_{r=0}^{r=r_{\rm cut}}
\frac{1}{2} [d(r^2 H) - r^2 dr H']\\
&=\pi v^2 r_{\rm cut}^2H(r_{\rm cut})
+
N_s,
\eeq
where $r_{\rm cut}$ is the IR cutoff length scale. As discussed in the main text, $r_{\rm cut}$ is chosen as the NGB wavelength, $r_{\rm cut} = k^{-1}$ with $k^{-1}\gg r_0$. 
For the profile $f(r)=2\arctan(r_0/r)$, $H(r) = 2r_0^2/(r^2 + r_0^2)$ and therefore 
$N_s\simeq 4\pi v^2 r_0^2 \ln [1/(kr_0)]$,
$N_a \simeq 2\pi v^2 r_0^2 +N_s$. Because $N_s$ is logarithmically enhanced, $N_a\simeq N_s$.

\section{Semi-local strings}
\label{app:semi-local}

Let us consider symmetry breaking $SU(2)_{\rm global}\times U(1)_{\rm gauge} \rightarrow U(1)_{\rm global}$ from the Lagrangian
\beq\label{eq:semi-local_Lagrangian}
\mathcal{L}
=
-\frac{1}{4}F^{\mu\nu}F_{\mu\nu}
+
|D_{\mu}\Phi|^2
-
\frac{1}{2}\lambda(|\Phi|^2-v^2)^2,
\eeq
where
\beq
\Phi = \begin{pmatrix}
    \Phi_1 \\
    \Phi_2
\end{pmatrix}
,~~~
D_{\mu} = \partial_{\mu} - ieA_{\mu}.
\eeq
For static field configurations, the energy per unit length of a string is 
\beq
\label{eq:semi-local_energy_0}
\mathcal{E}
=
\int d^2x \left[
|D_i\Phi|^2 + \frac{1}{2}B^2 + \frac{\lambda}{2}(|\Phi|^2-v^2)^2
\right].
\eeq
The Bogomol'nyi rearrangement gives
\begin{align}
\label{eq:semi-local_complete_square}
\mathcal{E}
&=
2\pi v^2 |n| 
+
\int d^2 x
\bigg\{
|(D_1 \pm iD_2)\Phi|^2 \\
&+ \frac{1}{2}[B\pm e(|\Phi|^2-v^2)]^2 
+
\frac{e^2}{2}(\beta-1)(|\Phi|^2-v^2)^2
\bigg\}, \nonumber
\end{align}
where we have defined 
\beq
\beta \equiv \frac{\lambda}{e^2},
\eeq
$n$ is the winding number, and the equation takes upper (lower) sign for positive (negative) $n$.

\subsection{$\beta=1$}
\label{app:semi-local_BPS}
For  $\beta = 1$, the Bogomol'nyi bound is satisfied when 
\beq\label{eq:semi-local_Bogomol'nyi_equations}
(D_1 \pm iD_2) \Phi = 0
,~~~
B\pm e (|\Phi|^2 - v^2) = 0.
\eeq
The equations take plus (minus) sign if the winding number $n$ is positive (negative).
To solve the equations, one can use the ansatz~\cite{Hindmarsh:1991jq}
\footnote{
More generally, one can use $f_2(r) {\rm e}^{im\theta}$ instead of $f_2(r)$ where $m$ is an integer. 
However, when discussing the stability of strings, it is enough to consider the case $m=0$~\cite{Hindmarsh:1991jq}.
}
\beq\label{eq:semi-local_ansatz}
\Phi_S = 
v\begin{pmatrix}
 f_1(r)e^{in\theta}   \\
 f_2(r)
\end{pmatrix}
,~~~
A_S(r) = 
\frac{1}{e}n\frac{a(r)}{r}.
\eeq
$f_1(r), f_2(r)$, and $a(r)$ are real functions of $r$. Note that if $f_2(r) = 0$, the ansatz is just that for the ANO strings in the Abelian-Higgs model. Substituting the ansatz into the Bogomol'nyi equations~\eqref{eq:semi-local_Bogomol'nyi_equations} gives, assuming unit winding $n=1$,
\beq\label{eq:semi-local_Bogomolyni_fg}
&f_1'(r) + \frac{a(r)-1}{r}f_1(r)=0,\\
&f_2'(r)+\frac{a(r)}{r}f_2(r) = 0, \\
&a'(r) + e^2 v^2 r[f_1^2(r)+f_2^2(r)-1] =0.
\eeq
Solving the first two equations gives
\beq
f_2(r) = Q \frac{f_1(r)}{r},
\eeq
with $Q$ being a (complex) constant. For arbitrary $Q$, the energy is the same, saturating the Bogomol'nyi bound, so $Q$ is a modulus.

Then, we can write Eq.~\eqref{eq:semi-local_ansatz} as~\cite{Achucarro:1999it}
\beq\label{eq:semi-local_ansatz2}
\Phi_S = 
v\frac{f_1(r)}{r}
\begin{pmatrix}
    r e^{i\theta}\\
    Q
\end{pmatrix}
=
v
\frac{1}{\sqrt{r^2+|Q|^2}}
\begin{pmatrix}
    r e^{i\theta}\\
    Q
\end{pmatrix}
{\rm e}^{u(r)/2},
\eeq
where 
\beq
u(r) \equiv \ln\left(\frac{|\Phi|^2}{v^2}\right)
,~~~
\frac{f_1(r)}{r} =\frac{e^{u(r)/2}}{\sqrt{r^2+|Q|^2}}.
\eeq
The first and third equations in Eq.~\eqref{eq:semi-local_Bogomolyni_fg} give 
\beq\label{eq:semi-local_u}
\nabla_r^2 u
+
2e^2v^2(1-e^{u}) = 
\frac{4|Q|^2}{(r^2+|Q|^2)^2},
\eeq
where $\nabla_r^2 = (1/r)d/dr(rd/dr)$. Because $|\Phi|\rightarrow v$, $u\rightarrow 0$ for $r\rightarrow \infty$. 
For large $r$, one then expands Eq.~\eqref{eq:semi-local_u} in terms of $u$ and obtains
\beq\label{eq:semi-local_u_large_r}
\nabla_r^2 u
-
2e^2v^2 u \simeq 
\frac{4|Q|^2}{(r^2+|Q|^2)^2}
,~~~
r\rightarrow \infty.
\eeq
If $u(r)$ follows a power law, then the second term on the left-hand side of Eq.~\eqref{eq:semi-local_u_large_r} dominates, and thus
\beq
u \simeq -\frac{1}{e^2 v^2}\frac{2|Q|^2}{(r^2+|Q|^2)^2}
,~~~
r\rightarrow \infty.
\eeq
Then, we get the asymptotic behavior of $f_1(r)$, $f_2(r)$ and $B(r)$ for $r\rightarrow \infty$,
\beq
f_1(r) \simeq &
\frac{r}{\sqrt{r^2+|Q|^2}}
\simeq
1-\frac{|Q|^2}{2r^2}
,\\
f_2(r) \simeq & \frac{Q}{r} - \frac{Q|Q|^2}{2r^3}, \\
B(r) = & ev^2(1-e^{u}) \simeq 
\frac{1}{e}
\frac{2|Q|^2}{(r^2+|Q|^2)^2},
\eeq

If $Q = 0$, the semi-local string profile is just the ANO string profile. If $|Q| \rightarrow \infty$, the solution of $u$ is trivially $u(r)=0$, and $\Phi_S$ in Eq.~\eqref{eq:semi-local_ansatz2} resides on the vacuum manifold for the whole space, which is known as a 2d skyrmion~\cite{Hindmarsh:1991jq}. Therefore, the $\beta=1$ semi-local-string solution interpolates between the ANO-string solution and skyrmion solution for arbitrary $Q$.

As in the skyrmion-line case in App.\,\ref{app:Skyrmion_charge}, the BPS semi-local string has a size and $U(1)$ modulus $Q$ whose moduli metric are logarithmically divergent; if $Q$ depends on $t$ or $z$, the string tension would logarithmically diverge.

To see the dynamics of the phase modulus,
let us consider another ansatz: $f_2(r) \to f_2(r) e^{i\omega t}$ and $A_0 = a_0(r)$, in addition to the ansatz in Eq.~\eqref{eq:semi-local_ansatz}. We require that $a_0(r)$ vanish at $r\rightarrow \infty$ for the energy to be finite. 
Then, the equation of motion for $f_2(r)$ is
\begin{align}
    f_2''(r) + \frac{1}{r}f_2'(r) &- e^2 v^2 f_2(r)(f_1^2(r)+f_2^2(r)-1)  \\
    &  - \frac{a^2}{r^2} f_2(r) + (\omega - e a_0(r))^2 f_2(r) = 0. \nonumber
\end{align}
For $r\rightarrow \infty$, $f_1(r) \to 1$, $f_2(r) \to 0$, $a(r) \to 1$, and $a_0(r) \to 0$, so the equation becomes
\begin{align}
    f_2''(r) + \frac{1}{r}f_2'(r) + \omega^2 f_2(r) = 0.
\end{align}
Note that this equation has the same structure as linearized Eq.~\eqref{eq:eom with charge} for the skyrmion-line case at large distances $r\gg \omega^{-1}$. The solutions of $f_2(r)$ are therefore combinations of Hankel functions $H_{0}^{(1,2)}(\omega r)$ with the $r^{-1/2}$ tail. The interpretation as incoming and outgoing waves and the resulting linear divergence of the energy per unit length are the same as in the skyrmion-line case.

\subsection{$\beta\neq 1$}

For $\beta\neq 1$, the Bogomol'nyi bound cannot be satisfied. In particular, for $\beta>1$, the semi-local string is unstable~\cite{Achucarro:1992hs,Hindmarsh:1991jq,Hindmarsh:1992yy}. The instability can be seen as follows. The energy of the semi-local string is given by  Eq.~\eqref{eq:semi-local_complete_square}. Since all terms in the integral are positive, the energy has a lower bound
\begin{align}
    \mathcal{E} \geq 2\pi v^2|n| + \Delta \mathcal{E},
\end{align}
where 
\begin{align}
    \Delta \mathcal{E} = \frac{e^2}{2}(\beta - 1)\int d^2 x \qty(|\Phi|^2 - v^2)^2.
\end{align}
The equality is saturated if and only if the Bogomol'nyi equations Eq.~\eqref{eq:semi-local_Bogomol'nyi_equations} are satisfied. The family of solutions to the Bogomol'nyi equations is parameterized by $Q$ in Eq.~\eqref{eq:semi-local_ansatz2}. For $\beta>1$, $\Delta \mathcal{E}$ is positive, and the energy is minimized when $|\Phi|^2 = v^2$ everywhere. As we have seen, this corresponds to the skyrmion solution, $Q \to \infty$. Therefore, at least, the lowest energy solution is unstable. Numerical calculations show that for $\beta > 1$, even a metastable solution does not exist, and the semi-local string is unstable~\cite{Achucarro:1992hs,Hindmarsh:1991jq,Hindmarsh:1992yy}. 
For $\beta<1$, on the other hand, $\Delta \mathcal{E}$ is negative, and such instability does not exist. In fact, the ANO string solution is stable for $\beta<1$~\cite{Achucarro:1992hs,Hindmarsh:1991jq,Hindmarsh:1992yy}.

\subsection{Coupling of semi-local strings to NGBs}
\label{app:semi-local coupling}
As in the skyrmion-line case, to derive the coupling of semi-local strings to NGBs, we consider the NGB modes on top of the semi-local string solution via
\begin{align}\label{eq:semi-local_NGB_fields}
    \Phi = U\Phi_S, \quad U \equiv \exp\qty(i \sum_{a=1,2} \sigma^a \frac{\Pi^a}{\sqrt{2} v}).
\end{align}
Substituting this into the Lagrangian~\eqref{eq:semi-local_Lagrangian}, we obtain the interaction terms between the semi-local string and NGBs,
\begin{align}
\label{eq:L_int_2}
\mathcal{L}_{\rm int}
=&
\frac{\sqrt{2}}{v}
(\partial_{\mu}\chi j^{\mu}_-
+
\partial_{\mu}\chi^* j^{\mu}_+
)
-\frac{i}{v^2}
(\chi^*\partial_{\mu}\chi - \chi\partial_{\mu}\chi^*)j_3^{\mu}\nonumber \\
&
+
\frac{|\Phi_S|^2}{v^2}\partial_{\mu}\chi^*\partial^{\mu}\chi,
\end{align}
where $\chi\equiv (\Pi_1+i\Pi_2)/\sqrt{2}$, $T_{\pm} = (\sigma_1 \pm i\sigma_2)/2\sqrt{2}$,
and
\beq
j^{\mu}_{\pm} &= 
i \left[
(D^{\mu}\Phi_{S})^{\dagger} T_{\pm}\Phi_{S}
-\Phi_{S}^{\dagger}T_{\pm}D^{\mu}\Phi_{S}
\right],
\\
j^{\mu}_{3} &=
i\left[
(D^{\mu}\Phi_{S})^{\dagger} T_{3}\Phi_{S}
-\Phi_{S}^{\dagger}T_{3}D^{\mu}\Phi_{S}
\right].
\eeq
Note that $j_{\pm}^{\mu}$ are the currents corresponding to the broken generators $T_{\pm}$. Since the Lagrangian~\eqref{eq:semi-local_Lagrangian} has the $SU(2)$ global symmetry, $\partial_{\mu}j^{\mu}_{\pm} = 0$. Consequently, with the boundary terms vanishing, the terms in the first bracket in Eq.~\eqref{eq:L_int_2} can be dropped, and $\mathcal{L}_{\rm int}$ becomes
\beq\label{eq:interaction_Lagrangian_3}
\mathcal{L}_{\rm int}
=
-\frac{i}{v^2}
(\chi^*\partial_{\mu}\chi - \chi\partial_{\mu}\chi^*)j_3^{\mu}
+
\frac{|\Phi_S|^2}{v^2}\partial_{\mu}\chi^*\partial^{\mu}\chi.
\eeq

\subsubsection{ANO case}
We first focus on the ANO branch,
$\Phi_S = (\phi_1,0)^T$. This includes the stable solution for $\lambda<e^2$ and the $Q=0$ member of the BPS family at $\lambda=e^2$.

Notice that the equation of motion for $A_{\mu}$ from the Lagrangian $\mathcal{L}$ gives
\beq
\partial_{\nu}F^{\nu\mu}_S
=ie[(D^{\mu } \Phi_S)^{\dagger}\Phi_S - \Phi_S^{\dagger}D^{\mu}\Phi_S]
=
2 e j_{3}^{\mu}.
\eeq
In the second step, we have used  $T_3\Phi_S = \frac{1}{2}\Phi_S$. Therefore, we can express $\mathcal{L}_{\rm int}$ as
\beq
\mathcal{L}_{\rm int}
&=
\frac{i}{v^2}\frac{1}{e}
F^{\nu\mu}_S
(\partial_{\nu}\chi^*\partial_{\mu}\chi)
+
\frac{|\Phi_S|^2}{v^2}\partial_{\mu}\chi^*\partial^{\mu}\chi.
\eeq
For a static straight long string, we have $F^{ij}_S = B_S\epsilon^{ij}$ ($i,j = 1,2$), and $\mathcal{L}_{\rm int}$ is simplified as
\beq
\mathcal{L}_{\rm int}
=
\frac{i}{v^2}\frac{1}{e}
B_S\epsilon^{ij}
(\partial_{i}\chi^*\partial_{j}\chi)
+
\frac{|\Phi_S|^2}{v^2}\partial_{\mu}\chi^*\partial^{\mu}\chi.
\eeq
Subtracting the free kinetic term $\partial_{\mu}\chi^*\partial^{\mu}\chi $, we have 
\beq\label{eq:L_int_Q=0}
\mathcal{L}_{\rm int}
=
\frac{i}{v^2}\frac{1}{e}
B_S\epsilon^{ij}
(\partial_{i}\chi^*\partial_{j}\chi)
+
\left(\frac{|\Phi_S|^2}{v^2} - 1
\right)
\partial_{\mu}\chi^*\partial^{\mu}\chi.
\eeq
Approximating the vortex as point-like source, the Lagrangian becomes
\beq\label{eq:semi-local_effective_point_source}
\mathcal{L}_{\rm eff}
\simeq 
-\left(
\frac{N_a}{v^2} i \epsilon^{ij}
\partial_{i}\chi^*\partial_{j}\chi
+
\frac{N_s}{v^2}
\partial_{\mu}\chi^*\partial^{\mu}\chi 
\right)
\delta^2(\boldsymbol{x}_{\perp} - \boldsymbol{X}(t)).
\eeq
where
\beq
&N_a \equiv -\int d^2 x \frac{B_S}{e}
,~~~
N_s \equiv -\int d^2 x (|\Phi_S|^2 - v^2).
\eeq

For the case $\lambda=e^2$ with $Q=0$,  $|N_a|=|N_s|$, which is guaranteed by the second equation of the Bogomol'nyi bound~\eqref{eq:semi-local_Bogomol'nyi_equations}, $|B_S/e|= |(|\Phi_S|^2 - v^2)|$. For the case $\lambda < e^2$, the string vortex is ANO, and we know for large distance $r$, the scalar part $f_1(r) \rightarrow 1- \exp{(-\sqrt{\lambda} v r )}$, so the size of the ANO core (Higgs sector) is thus $r_H\sim (1/\sqrt{\lambda}) v^{-1}$. Therefore
\beq
N_s = -\int dx^2
\left(|\Phi_S|^2 - v^2
\right)
\sim 
r_H^2 v^2
\sim 
\frac{1}{\lambda }.
\eeq
The antisymmetric part can be precisely computed,
\beq
N_a =- \int dx^2\frac{1}{e }B_S 
=
-\frac{2\pi n}{e^2}.
\eeq
Therefore, we have
\beq
\frac{N_s}{N_a}
\sim -\frac{e^2}{\lambda} \frac{1}{n}
,~~~
\text{for $\lambda \leq e^2$}.
\eeq

\subsubsection{$Q\neq 0$ case}

The general BPS solution for the case $\lambda=e^2$ is Eq.~\eqref{eq:semi-local_ansatz2}, where the second component of $\Phi_S$ also depends on $r$. We still have $\Phi_S \rightarrow (v{\rm e}^{i\theta}, 0)^T$ as $r\rightarrow \infty$, so $\Pi$ fields in Eq.~\eqref{eq:semi-local_NGB_fields} still represent NGBs. 
The discussion about $j_{\pm}^{\mu}$ remains unchanged, $\partial_{\mu}j_{\pm}^{\mu}=0$, and we can still drop the first term in Eq.~\eqref{eq:L_int_2}. But now $j_3^{\mu}$ becomes different, for we no longer have $T_3\Phi_S = \frac{1}{2}\Phi_S$.

Next, we apply the Bogomol'nyi bound \eqref{eq:semi-local_Bogomol'nyi_equations}. 
Without the loss of generality, we assume that the winding number $n$ is positive, so we have $(D_1 + iD_2) \Phi = 0$, which gives
\beq\label{eq:Bogomol'nyi_bound_2}
D_i\Phi_S = -i\epsilon_{ij}D_j\Phi_S
,~~~
i,j=1,2.
\eeq
For a nearly static long straight string, we have $\partial_{0,z} = 0$ and $A^S_{0,z} = 0$. Therefore, we have $j_3^{0,z} = 0$, and the remaining components are
\beq
j^{i}_{3} &=
i\left[
(D^{i}\Phi_{S})^{\dagger} T_{3}\Phi_{S}
-\Phi_{S}^{\dagger}T_{3}D^{i}\Phi_{S}
\right]
,~
i=1,2.
\eeq
Then, applying the Bogomol'nyi bound Eq.~\eqref{eq:Bogomol'nyi_bound_2}, we can rewrite $j_3^{i}$ as
\beq\label{eq:j_3-2}
j^{i}_{3} 
&=
-\epsilon^{ij}\partial^j
(
\Phi_{S}^{\dagger}T_{3}\Phi_{S}
)
=
-
\frac{\epsilon^{ij}}{2}\partial^j
(|\phi_1|^2 - |\phi_2|^2).
\eeq
As $r\rightarrow \infty$, we have $|\phi_1|^2\rightarrow v^2$ and $|\phi_2|^2\rightarrow 0$ and thus $(|\phi_1|^2 - |\phi_2|^2)\rightarrow v^2$. Motivated by this, we subtract a $v^2$, which will not change $j_3$ since $\partial^j(v^2) = 0$, 
\beq
j_3^i
&=
-
\frac{\epsilon^{ij}}{2}\partial^j
\left(|\phi_1|^2 - |\phi_2|^2 -v^2\right).
\eeq
Now the boundary term vanishes, so we can drop the total-derivative in the Lagrangian.
Consequently, $\mathcal{L}_{\rm int}$ in Eq.~\eqref{eq:interaction_Lagrangian_3} becomes (with the free kinetic term $\partial_{\mu}\chi\partial^{\mu}\chi^*$ subtracted)
\beq\label{eq:L_int_final}
\mathcal{L}_{\rm int}
=
&
-i
\left(\frac{|\phi_1|^2-|\phi_2|^2}{v^2}-1\right)
\epsilon^{ij}
\partial_i\chi^*\partial_{j}\chi \\
&+
\left(
\frac{|\Phi_S|^2}{v^2}
-1
\right)
\partial_{\mu}\chi^*\partial^{\mu}\chi.
\eeq
If we take $Q=0$ and thus $\phi_2 = 0$, it recovers the previous result Eq.~\eqref{eq:L_int_Q=0}, as expected.

Again, if we can approximate the vortex as point-like source, the effective Lagrangian becomes Eq.~\eqref{eq:semi-local_effective_point_source}, but now with the following normalization factors
\beq
&N_a \equiv \int d^2 x \left[\left(|\phi_1|^2-|\phi_2|^2-v^2\right)\right],\\
&N_s \equiv -\int d^2 x (|\Phi_S|^2 - v^2).
\eeq
Furthermore, we know that as $r\rightarrow \infty$,
\beq
&
\frac{|\phi_1|^2-|\phi_2|^2}{v^2}-1
=
{\rm e}^{u(r)}
\frac{r^2 - |Q|^2}{r^2+|Q|^2}
-1 \\
&
\rightarrow
-
\frac{2|Q|^2}{r^2}
+ \mathcal{O}(r^{-4}),
\eeq
and
\beq
&
\frac{|\Phi_S|^2}{v^2}
-1
=
{\rm e}^{u(r)} - 1
\rightarrow 
- \frac{2}{e^2v^2}\frac{|Q|^2}{r^4}
+ \mathcal{O}(r^{-6})
\eeq
Therefore, $N_a$ has an IR logarithmic divergence like that in the skyrmion-line case, while $N_s$ is not divergent. Furthermore, $N_s$ can be precisely calculated with the Bogomol'nyi bound applied, $N_s = \frac{2\pi n }{e^2}$, which is independent of $Q$.

\section{Computation of the NGB emission rate}
\label{app:emission rate}
In this appendix, we compute the NGB emission rate from strings. We take $\boldsymbol{X}(t,z) = X_0 {\rm cos}(\omega t) \hat{n}_x$.
For both skyrmion lines and semi-local strings, the interactions between the topological line and NGBs contain a symmetric term and an antisymmetric term. We first focus on the symmetric term.

The matrix element for the pair-production of $\chi$ and $\chi^*$ with four-momenta $k_1$ and $k_2$ is
\begin{align}
    {\cal M} =& \frac{N_s}{v^2} k_1\cdot k_2 \int dz e^{-i(k_{1,z} + k_{2,z})z} \times F, \nonumber \\
    F \equiv & \int dt e^{- i ({\bf k}_1 + {\bf k}_2)_\perp \cdot {\bf X}(t)}e^{i (\omega_1 + \omega_2) t}.
\end{align}
For $\boldsymbol{X}(t) = X_0 {\rm cos} \omega t ~\hat{x}$, assuming $X_0|k_{1,x} + k_{2,x}| \ll 1$,
\begin{align}
\label{eq:F_app}
    F \simeq - \frac{i}{2}X_0(k_{1,x}+ k_{2,x}) \int dt e^{i(\omega_1 + \omega_2 -\omega)}.
\end{align}
In taking the square of ${\cal M}$, we use
\begin{align}
    \left|\int dz e^{-i(k_{1,z} + k_{2,z})z}\right|^2 = 2 \pi \delta(k_{1,z}+ k_{2,z})L, \nonumber \\
    \left| \int dt e^{i(\omega_1 + \omega_2 -\omega)}\right|^2= 2\pi \delta (\omega_1 + \omega_2 -\omega) T
\end{align}
where $L$ is the total length of the string and $T$ is the total time.

The probability to emit the NGBs per time and per length is
\begin{align}
    \frac{d^2p}{dt dL} =& \frac{1}{LT} |{\cal M}|^2 \widetilde{dk_1} \widetilde{dk_2}  \\
    =& (2\pi)^2\frac{N_s^2 X_0^2}{4 v^4} (k_1\cdot k_2)^2 (k_{1,x}+ k_{2,x})^2 \nonumber \\ &\times\delta(k_{1,z}+ k_{2,z}) \delta (\omega_1 + \omega_2 -\omega)  \widetilde{dk_1} \widetilde{dk_2}, \nonumber 
\end{align}
where
\begin{equation}
    \widetilde{dk_i}  = \frac{d^3 k_i}{(2\pi)^3 2 \omega_i}.
\end{equation}
The energy of NGBs emitted from the string per length and per time is
\begin{align}\label{eq:NGB_emission_skyrmion}
    \frac{d^2E}{dt dL} = & \int \widetilde{dk_1} \widetilde{dk_2} (\omega_1  + \omega_2) (2\pi)^2\frac{N_s^2 X_0^2}{4 v^4} (k_1\cdot k_2)^2  \\ &\times (k_{1,x}+ k_{2,x})^2\delta(k_{1,z}+ k_{2,z}) \delta (\omega_1 + \omega_2 -\omega). \nonumber 
\end{align}

For $m_\chi >\omega/2$, the energy conservation cannot be satisfied and the emission rate vanishes.
We perform the integration for $m_\chi \ll \omega/2$.
With 
\begin{equation}
P^{\mu} \equiv k_{1}^{\mu} + k_2^{\mu}
,
\end{equation}
the integral can be rewritten as
\begin{equation}
I = C \cdot \int \widetilde{dk_1} \widetilde{dk_2}~
P^0 \cdot \frac{(P^2)^2}{4} \cdot P_x^2 
\cdot \delta(P_z)\delta(P^0-\omega).
\end{equation}
where $C \equiv (2\pi)^2 \frac{N_s^2 X_0^2}{4v^4}$. Inserting the identity $1=\int d^4P \delta^{(4)}(P-k_1-k_2)$ into the integral,
\begin{align}
I =& C\cdot \int d^4 P ~P^0 \cdot \frac{(P^2)^2}{4} \cdot P_x^2 
\cdot \delta(P_z)\delta(P^0-\omega)\\
&\times
\int \widetilde{dk_1} \widetilde{dk_2}~\delta^{(4)}(P-k_1-k_2).\nonumber
\end{align}
The second line is Lorentz invariant and can be evaluated in the rest frame $P^{\mu} = (P^0, \boldsymbol{0})$,
\begin{align}
\int \widetilde{dk_1} \widetilde{dk_2}~\delta^{(4)}(P-k_1-k_2) 
 =
\frac{1}{128 \pi^5}. 
\end{align}
Therefore, the full integral is
\begin{align}
I &= \frac{C}{128 \pi^5}
\int d^4 P ~P^0 \cdot \frac{(P^2)^2}{4} \cdot P_x^2 
\cdot \delta(P_z)\delta(P^0-\omega) \nonumber \\
&=
\frac{C}{128 \pi^5}\int dP_x dP_y ~\omega \frac{(\omega^2 - P_x^2 - P_y^2)^2}{4} P_x^2.
\end{align}
Because $P^2 = 2k_1 k_2 = 2(\omega_1\omega_2 - \omega_1\omega_2\cos\theta)\geq 0$, the integration range is $P_x^2+P_y^2 \leq \omega^2$. 
Therefore
\begin{align}
\label{eq:integral_emission_skyrmion}
I&=
\frac{C}{128 \pi^5}\int_{P_x^2+P_y^2 \leq \omega^2} dP_x dP_y ~\omega \frac{(\omega^2 - P_x^2 - P_y^2)^2}{4} P_x^2 \nonumber \\
& = 
\frac{C}{128 \pi^5}
\int_0^{2\pi}d\varphi \int_0^{\omega} dr ~r \cdot \omega \frac{(\omega^2 - r^2)^2}{4} (r\cos\varphi)^2 \nonumber\\
& =\frac{C}{128 \pi^5} \times \frac{\pi}{96} \omega^9 .
\end{align}

Next, we focus on the antisymmetric interaction term $\sim \epsilon^{ij}\partial_i \chi^* \partial_j\chi$. 
To calculate the corresponding NGB emission rate, we just need to replace $k_1\cdot k_2$ by $\epsilon^{ij}k_{1,i}k_{2,j}
=
k_{1,x}k_{2,y}- k_{1,y}k_{2,x}$ in the computation above. The emission rate in Eq.~\eqref{eq:NGB_emission_skyrmion} now becomes
\beq
    \frac{d^2E}{dt dL} &= C_a \int \widetilde{dk_1} \widetilde{dk_2} (\omega_1  + \omega_2) (\epsilon^{ij}k_{1,i}k_{2,j})^2  \\ &~~\times (k_{1,x}+ k_{2,x})^2\delta(k_{1,z}+ k_{2,z}) \delta (\omega_1 + \omega_2 -\omega) \\
    &=
    C_a \int d^4 P P^0 P_x^2 \delta(P_z) \delta(P^0-\omega) \\
    &~\times
    \int \widetilde{dk_1} \widetilde{dk_2} \delta^{(4)}(P-k_1-k_2) (k_{1,x}P_y - k_{1,y}P_x)^2
\eeq
where $C_a \equiv (2\pi)^2 \frac{N_a^2 X_0^2}{4v^4}$. To evaluate the phase-space integral in the second line, we define the following Lorentz tensor,
\beq
T^{\mu\nu}\equiv
\int \widetilde{dk_1} \widetilde{dk_2} \delta^{(4)}(P-k_1-k_2) k_1^{\mu}k_1^{\nu}.
\eeq
Since the only external four-vector for $T^{\mu\nu}$ is $P^{\mu}$, we can in general express it as 
\beq
T^{\mu\nu} = a(s) g^{\mu\nu} + b(s)P^{\mu}P^{\nu}
,~~~
s\equiv P^2.
\eeq
One can check that $g_{\mu\nu}T^{\mu\nu} = \int \widetilde{dk_1} \widetilde{dk_2} \delta^{(4)}(P-k_1-k_2) k_1^2 = 0$ since $k_1^2 = 0$ in the massless limit, which gives $4a+bs = 0$, and $P_{\mu}P_{\nu}T^{\mu\nu} = \int \widetilde{dk_1} \widetilde{dk_2} \delta^{(4)}(P-k_1-k_2) (k_1\cdot P)^2 = \frac{s^2}{4}\cdot\frac{1}{128\pi^5}$ since $k_1\cdot P = s/2$, which gives $as+bs^2 = \frac{s^2}{4}\cdot\frac{1}{128\pi^5}$. Consequently, one has $a = -\frac{s}{12}\cdot\frac{1}{128\pi^5}$ and $b = \frac{1}{3}\cdot\frac{1}{128\pi^5}$. Then,
\beq
&\int \widetilde{dk_1} \widetilde{dk_2} \delta^{(4)}(P-k_1-k_2) (k_{1,x}P_y - k_{1,y}P_x)^2 \\
&= P_y^2 T^{xx}
+
P_x^2 T^{yy}
-
2P_xP_y T^{xy} \\
& = \frac{1}{128\pi^5} \frac{P^2(P_x^2+P_y^2)}{12}.
\eeq
Furthermore,
\beq\label{eq:integral_emission_antisymmetric}
\frac{d^2E}{dtdL}
&=
\frac{C_a}{128\pi^5}
\int_{P_x^2+P_y^2\leq \omega^2}
dP_xdP_y \\
&
\left[\omega P_x^2 
\frac{(\omega^2-P_x^2-P_y^2)(P_x^2+P_y^2)}{12}
\right]\\
& =
\frac{C_a}{128\pi^5} \times 
\frac{\pi}{288} \omega^9.
\eeq
Therefore, combining the emission rates from symmetric and antisymmetric interactions, Eqs.~\eqref{eq:integral_emission_skyrmion} and \eqref{eq:integral_emission_antisymmetric}, we have 
\begin{align}
\label{eq:emission rate}
    \frac{d^2E}{dt dL} = \begin{cases}
    \frac{ (N_s^2 + N_a^2/3) X_0^2 \omega^9}{12288 \pi^2 v^4} 
    &: m_\chi \ll \omega/2 \\
      0 &: m_\chi > \omega/2
    \end{cases}.
\end{align}
Note that the vanishing rate for $m_\chi >\omega/2$ is due to the approximation in Eq.~\eqref{eq:F_app}. Once the terms higher order in $X_0$ are included, the emission rate is, although suppressed, non-zero even for $m_\chi >\omega/2$.

To compute the emission rate, we used the approximation $X_0|k_{1,x} + k_{2,x}| \ll 1$ in Eq.~\eqref{eq:F_app}, which is valid if $X_0\omega \ll 1 $. For generic $X_0$, the emission rate is given by Eq.~\eqref{eq:emission rate} times $g(X_0\omega)$, where $g$ is a function with $g(x)\simeq 1$ for $x\ll 1$.
To analyze NGB emission from string loops, we will extrapolate our computation to $X_0 \omega \sim 1$, for which the rate is expected to be on the order of Eq.~\eqref{eq:emission rate} with $X_0\sim \omega^{-1}$.

Note that $\boldsymbol{X}(t,z) = X_0 {\rm cos}(\omega t) \hat{n}_x$ is not a solution of the equation of motion of free strings. The estimated rate can be understood as that from an approximately straight part of the string network that is forced to oscillate by other parts of the network.

For $\boldsymbol{X}(t,z) = X_0 {\rm cos}(\omega (t-z)) \hat{n}_x$, which is a solution of the equation motion for free strings, we find that the emission rate vanishes to all orders in $X_0$. Technically, this is because $P^0=|P_z|=n\omega$ is enforced for $n$th order in $X_0$, and the phase space for $P_{x,y}$ vanishes. Physically, such $\boldsymbol{X}(t,z)$ corresponds to upward-moving position moduli, which cannot decay into two massless particles. 
For $\boldsymbol{X}(t,z) = X_0 {\rm cos}(\omega t) {\rm cos}(\omega z) \hat{n}_x$, we find that the contribution from the leading order in $X_0$ vanishes. This is because the leading order term corresponds to the decay of a position moduli into two NGBs, which is kinematically forbidden. A non-zero rate arises from $O(X_0^2)$ in the matrix element, which corresponds to the collision of an upward-moving modulus with a downward-moving modulus.

\bibliography{references}
\end{document}